# On Apparent Superluminal Motion in Astrophysical Jet Models

Ernst Karl Kunst

**Astrophysical models developed to explain superluminal motion in beaming phenomena are shown to be incomplete and the correct kinematical expression in any case to predict projected velocities slower than the speed of light. The observed superluminal motions in the plane of the sky are proposed to be real.**

Radio astronomers have known of superluminal velocities in extragalactic radio sources for now more than 30 years (in the optical region since some time). The phenomenon has been widely discussed and reviewed ever since [1], [2]. The evidence is that rapid superluminal motion is a very common feature among celestial sources, even in the Milky Way galaxy. From the competing ingenious models that have been developed to explain the phenomenon within the framework of relativistic kinematics the one first proposed by Rees has been accepted as the most attractive theoretical explanation [3], [4]. It involves relativistic motion of the radio (or light) emitting regions in the superluminal sources. The superluminal effect thus arises from the time delay between the two components, when the angle ϕ of the motion to the line of sight is small. The relativistic motion towards the observer has the additional effect of enhancing the flux density (Doppler boosting).

If $v_{app}$ is the apparent velocity observed by the observer at the distant point O and v the space velocity of some directly observed feature, then $v_{app}$ and v are apparently related by the simple formula (see Fig. 2a)

$$v_{app} = \frac{v \sin\phi}{1 - \frac{v}{c}\cos\phi}. \qquad (1)$$

It is clear that by inserting appropriate values of v and ϕ into equation (1) in many cases superluminal velocity results and, thus, this phenomenon apparently can be shown to be an illusionory effect. But it seems that the physial notion underlying the formula above is much too simple and needs some modification, which eventually will lead always to projected velocities $v_{app} = v \sin\phi < c$.

Let us reanalyse the problem from a purely (special) relativistic and kinematical point of view. For this purpose consider the projection of a "jet" in the plane of the sky to be caused (in first approximation) by a more or less stable structure of length r between the fixed points A (begin) and B (end). In the kinematical models, which lead to formula (1), the superluminal phenomenon is presumed to represent a physical motion from point A to point B almost directly towards the distant observer at O (Fig. 2a). This implies the point A (begin of the jet) to be at rest with respect to the observer resting at the point O at rest (in view of the involved high relativistic expansion velocities v → c appears any relative motion between the points A and O negligible, as well as effects of gravity and rotation).

On the other hand, there always can be introduced an oppositely to r directed counter jet pointing from point A at rest to a new point C so that CAB = 2r. In the over-simplified geometrical picture leading to equation (1) would the apparent motion of a physical feature from point A to point C be described by

$$v'_{app} = \frac{v\sin\phi}{1 + \frac{v}{c}\cos\phi}, \qquad (2)$$

delivering always $v'_{app} < v\sin\phi < v_{app}$.

Let CAB be normal to the line of sight of the observer at O. Light signals, which are emitted simultaneously from point A at the very center of this structure towards the points B and C according to the Einsteinian simultaneity definition in reference [5] and [6], would also arrive there simultaneously after time Δt = r/c, covering in this time distance 2r (see Fig. 1).

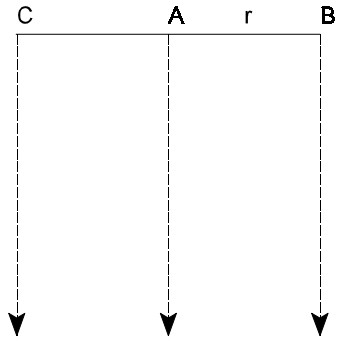

Fig. 1

If the distance to the observer at the point O is denoted d, then the velocity of the oppositely directed light signals (or particles travelling with about the speed of light) can due to their simultaneous arrival at the points B and C with respect to the observer at O be expressed as [(2d + 2r) - 2d]/(2Δt) = c. It seems that this "natural" projection of an extended structure in the plane of the sky on the grounds of the validity of the Einsteinian simultaneity definition, where light signals need the least possible time Δt to measure it from one end to the other so that 2r = 2cΔt be in any case the basic requirement for the comparison of measurements. This must also be valid even if only one jet seems to exist, because, as already mentioned, in any such case a counter jet can always be introduced.

Suppose a light signal flashing simultaneously over the whole length CAB = 2r towards the observer who happens to rest at point O. It is clear that also the light rays from the emitters B and C arrive simultaneously at O and, thus, no time lag between the arrival times can be observed.

If now CAB = 2r is inclined to the line of sight at angle ϕ (which is the only interesting case), with point B being the near end with respect to O, then a simultaneous flash of light towards O would evidently not arrive simultaneously at the observer resting there. Rather the light from the emitter B would arrive first and the light from emitter C last, lagging behind time 2r cosϕ/c, so that the observer at point O will observe the light flash moving apparently from B to C with velocity c sinϕ/cosϕ in this time (see Fig. 2b).

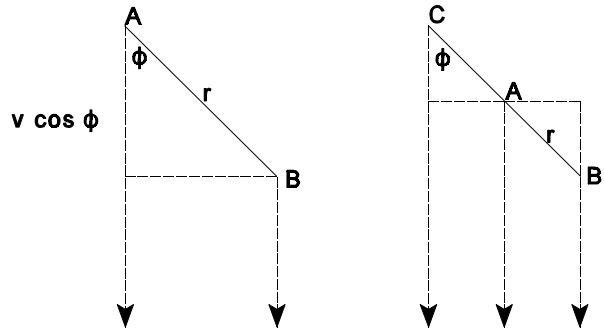

Fig. 2a          O                Fig. 2b          O

It is clear that this physical picture also is valid for the "natural" projection above of light signals (or particles with about this speed) arriving simultaneously at the points B and C. The light from emitter C also lags behind the light from emitter B by time 2r cosϕ/c, as it generally does from any point at distance r' < r between the points A and C by time 2r' cosϕ/c, compared with the light signal from the respective (counter) point at distance r' between points A and B. Hence, the motion of physical features moving simultaneously with or with about the speed of light from the center point A to the points B and C, respectively, must be delayed in the direction of point B and sped up in the opposite direction towards point C by halve this time, i. e. Δt cosϕ. This is evidently valid for any physical motion starting simultaneously from the point A towards the points B and C. These effects cancel exactly and, thus, compensate the apparent shortening of time by -v cosϕ/c in the denominator of the right-hand side of (1) and the delay of the same order of magnitude in (2) so that these equations have to be corrected to

$$V_{app} = V'_{app} = \frac{v\sin\phi}{1 - \frac{v\cos\phi}{c} + \frac{v\cos\phi}{c}} = \frac{v\sin\phi}{1 + \frac{v\cos\phi}{c} - \frac{v\cos\phi}{c}} = v\sin\phi. \qquad (3)$$

The result (3) is also backed by considering that the points A and O are resting one to another and, thus, the time interval $\Delta t_A$ of a physical motion or two counter directed ones with respect to point A must in any case be measured in both frames of reference alike, i. e. $\Delta t_O = \Delta t_A$. The previous expression becomes in the light of the foregoing to $[(d + r \cos\phi + r \sin\phi) + (d - r \cos\phi + r \sin\phi) - 2d]/(2\Delta t) = v \sin\phi$ (see Fig. 2b). It is clear that this reflects a fundamental property of space-time symmetry and, therefore, is totally independent of the accidental existence or visibleness of a counter jet or a counter directed motion to any moving physical feature in the plane of the sky.

Our analysis leads to the conclusion that on the grounds of Einsteinian relativistic kinematics superluminal motion in beaming phenomena at whatever angle to the line of sight of an observer at Earth cannot occur. Hence the observed superluminal motions among celestial objects must be of other origin than proposed by the hitherto discussed astrophysical models. These all have in common that the axiom of special relativity, that no real physical motion can exeed the speed of light, even holds if vast cosmological distances are involved and, therefore, contrary observations must be illusionory effects. But if formula (3) correctly describes every projected velocity v in the framework of special relativity, even if $v \to c$ and angle $\phi \to 0$, then the conclusion is unavoidable that the observed superluminal motions in the sky are real. This has already been proposed as a quintessence of a distance-dependent Lorentz transformation [7].